# Nonreciprocal Scattering by *PT*- symmetric stack of the layers

*O.V. Shramkova\*, Senior Member, IEEE, G.P. Tsironis*
*Crete Center for Quantum Complexity and Nanotechnology, University of Crete*
*P.O. Box 2208, 71003 Heraklion, Greece*
*Tel: (30) 2810 39 4322, e-mail*: oksana@physics.uoc.gr

**ABSTRACT**
The nonreciprocal wave propagation in *PT*-symmetric periodic stack of binary dielectric layers characterised by balances loss and gain is analysed. The main mechanisms and resonant properties of the scattered plane waves are illustrated by the simulation results, and the effects of the periodicity and individual layer parameters on the stack nonreciprocal response are discussed. Gaussian beam dynamics in this type of structure is examined. The beam splitting in *PT*-symmetric periodic structure is observed. It is demonstrated that for slant beam incidence the break of the symmetry of field distribution takes place.
**Keywords**: parity-time symmetry, periodic structure, anisotropic transmission resonance, coherent perfect absorber laser.

## 1. INTRODUCTION

Optical metamaterials have been extensively studied during the last decade. Nonreciprocity in artificial materials has recently been a subject of particular interest in the context of their applications in microwave and photonic devices. The discovery of parity-time (*PT*)-symmetric media have shown significant potential for advancing toward the goal of new metamaterial design. The *PT*-related concepts can be realized in artificial optical materials that rely on balanced gain and loss regions. In this framework, *PT*-symmetry demands that the complex refractive index obeys the condition $n(\vec{r}) = n^*(-\vec{r})$. In the context of optics, *PT*-symmetric systems have demonstrated several exotic features, including unidirectional invisibility [1], coherent perfect absorption [2], nonreciprocity of light propagation [3][4], beam refraction [3] and various extraordinary nonlinear effects [5].

Since majority of the currently available metamaterials are fabricated as stacked layers, it is expedient to investigate the scattering characteristics in the planar layered *PT*-symmetric structures. Such an approach provides insight in the fundamental mechanisms of phase transitions and of resonant phenomena. The existence of transmission resonances in which the reflectance vanishes only for waves incident from one side of the structure and transmission coefficient is equal to unity, which we refer to as anisotropic transmission resonances (ATRs), in the spectrum of 1D *PT*-symmetric photonic heterostructure were recently discussed in [6]. It was demonstrated in [7] that optical medium, consisting of a uniform index grating with two homogeneous and symmetric gain and loss regions, can behave simultaneously as a laser oscillator, emitting outgoing coherent waves, and as a coherent perfect absorber (CPA), absorbing incoming coherent waves.

The optical properties of *PT*-symmetric periodic stack of the layers with balanced loss and gain have been examined in the work. The resonant phenomena and effect of structure periodicity on the reflectivity and transmittivity of *PT*-symmetric stack are analysed. Gaussian beam dynamics in this periodic structure is examined.

## 2. PROBLEM STATEMENT

A periodic structure shown in Fig. 1 is composed of the stacked binary dielectric layers of identical thickness *d*, with complex-conjugate dielectric permittivities $\varepsilon = \varepsilon' - i\varepsilon''$ and $\varepsilon^* = \varepsilon' + i\varepsilon''$ ($\varepsilon'$ and $\varepsilon''$ are positive) corresponding to balanced gain and loss regions, respectively. The layers are assumed isotropic and of infinite extent in the *x*- and *y*-directions. The stack has total thickness $L = 2Nd$, where *N* is the number of periods, and is surrounded by linear homogeneous medium with dielectric permittivity $\varepsilon_a$ at $z \leq -L/2$ and $z \geq L/2$. We are interested in studying the electromagnetic response under oblique TM-wave illumination.

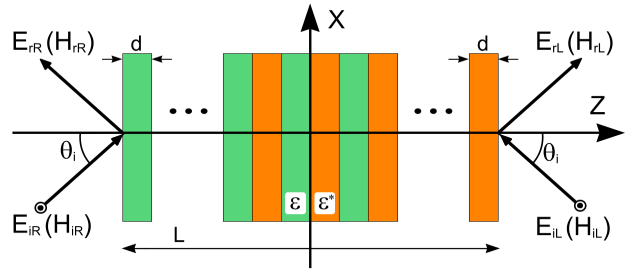

*Figure 1. Geometry of the problem.*

To evaluate scattering properties and symmetry breaking transitions in *PT*- symmetric periodic structure, we use the scattering matrix (S-matrix) formalism. In the approximation of non-depleting waves, this is accomplished with the aid of the transfer matrix of whole stack composed by the transfer matrices of individual layers. In our work we use two alternative definitions of the scattering matrix for the determination unidirection invisibility points, CPA laser threshold points, where we can observe pole and zero of S-matrixes, and measurement of *PT*-symmetry breaking. The analyse of the expressions for reflection coefficients has enabled an insight in the fundamental properties of the scattering by periodic stack of the layers illuminated by obliquely incident plane wave. As the result of analytical and numerical investigation we demonstrate that tunneling phenomenon in periodic structures is connected with excitation of surface waves at the boundaries separating gain and loss regions within each unit cell and tunneling conditions for periodic stack can be reduced to the conditions for one period. Alternatively, it is shown that coherent perfect absorber (CPA) laser states are mediated by excitation of surface modes localised at all internal boundaries of the structure.

### 3. PLANE WAVE SCATTERING BY *PT*-SYMMETRIC STACK

To illustrate the effect of the stack periodicity, the reflectivity and transmittivity for both left and right normal incidence of TM plane waves are displayed in Fig. 2. The constituent layers have identical thicknesses $d$=125 µm and $\varepsilon' = \varepsilon'' = 0.1$. The media surrounding the stack have permittivity $\varepsilon_a = 1$. The considerable differences in reflectivity/transmittivity of *PT*-symmetric bilayer (Figs.2a) and periodic stack with *N*=10 (Figs.2b) are evident, especially at low frequencies, and connected with Bragg resonances for thicker structure. It is evident that at opposite incidence the reflectivity exhibits noticeable nonreciprocity and we can observe 2 ATRs for right incidence and wave amplification for the left one. It is necessary to note, that the frequencies of ATRs do not depend on the number of structure periods. At the same time the magnitudes of the reflectivity/transmittivity will be changed close to the points of ATR. In Fig.2 these resonances are marked by vertical dotted lines. Peaks of reflectivity and transmittivity at $\omega = 1.453 \times 10^{13}$ $s^{-1}$ for bilayer structure and multiple peaks corresponding to exceptional points of the scattering matrix eigenvalues for *N*=10 are the CPA laser resonant points. All other reflectionless transmission resonances (Bragg resonances) in Fig.2b are attributed to the increase of the stack overall thickness. At these resonance points structure reflectionless for both directions of incidence. The amplification of reflected waves is connected with a constructive interaction between the forward- and backward-propagating waves.

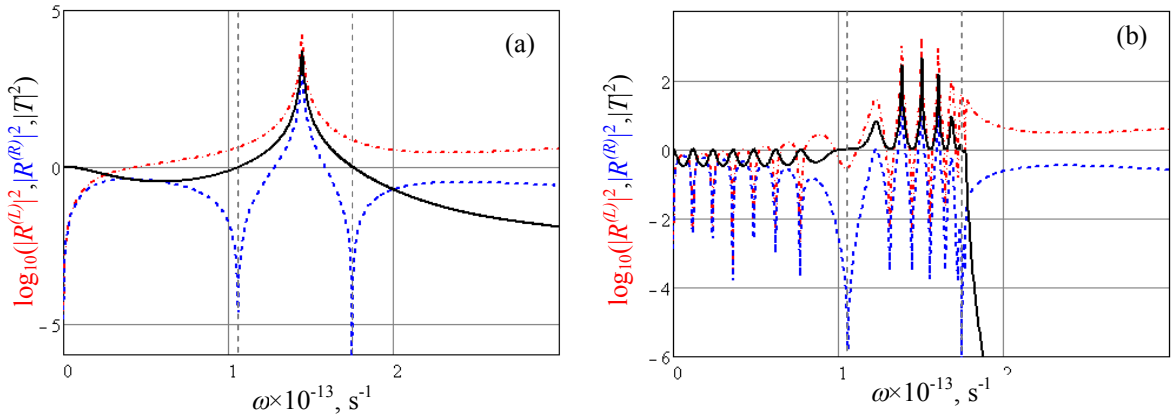

*Figure 2. TM wave reflectance ($\left|R^{(L)}\right|^2$ – dash-dot red curve, $\left|R^{(R)}\right|^2$ – dashed blue curve) and transmittance (solid black curve), for PT-symmetric stack with $\varepsilon' = 0.1$, $\varepsilon'' = 0.1$, d=125 µm, (a) - N=1, (b) - N=10 at the normal incidence.*

To gain insight in the effect of the incidence angle on the scattering properties of *PT*-symmetric stacks, TM-wave reflectance and transmittance were simulated at variable $\theta_i$. It was demonstrated that as a result of incidence angle increasing the symmetry breaking transition moves to the area of higher frequencies. At higher $\theta_i$ we can get only one ATR resonance for left incidence in the *PT*-symmetry broken area and one for right incidence at frequencies corresponding to *PT*-symmetrical structures.

### 4. GAUSSIAN BEAM SCATTERING BY *PT*-SYMMETRIC STRUCTURES

To investigate the Gaussian beam evolution in *PT*-symmetric stacks the magnetic field distribution for the binary dielectric layers with the default parameters as in Fig. 2(a) at frequency $\omega = 1.761 \times 10^{13}$ $s^{-1}$ close to ATR is displayed in Fig. 3. In our calculations the spot radius of the beam was equal to 0.3mm. Comparison of Figs. 3(a) and 3(b) for left and right normal incidence demonstrates strong nonreciprocal response of the structure. It can be seen that the field is localised at the layer interfaces. The strong amplification of reflected beam at left incidence and high transmission of incident beam for right incidence are observed.

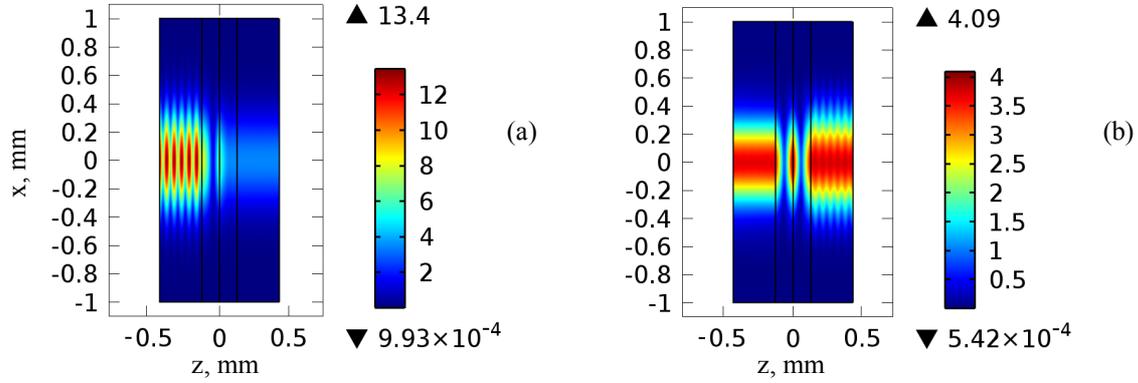

*Figure 3. Distribution of the magnetic-field magnitude $|H_y|$ (the units of measurement are [A/m]) for Gaussian beam incidence from left (a) and right (b) at $\theta_i = 0°$, $\varepsilon' = 0.1$, $\varepsilon'' = 0.1$, $d=125\ \mu m$, $N=1$, $\omega = 1.761\times10^{13}\ s^{-1}$ (The numerical simulation was carried out by package COMSOL Multiphysics).*

Fig.4 shows magnetic field distribution during Gaussian beam propagation through the *PT*-symmetric periodic stack of dielectric layers with $N=10$ at frequencies $\omega = 2.833\times10^{13}\ s^{-1}$ (Figs.4(a,b)) close to ATR and $\omega = 2.417\times10^{13}\ s^{-1}$ (Figs.4(c,d)) close to CPA lasing point. The beam splitting and shifting with increase of period numbers are observed. The amplification and widening of reflected and transmitted beams at frequency of CPA lasing for both directions of incidence are demonstrated.

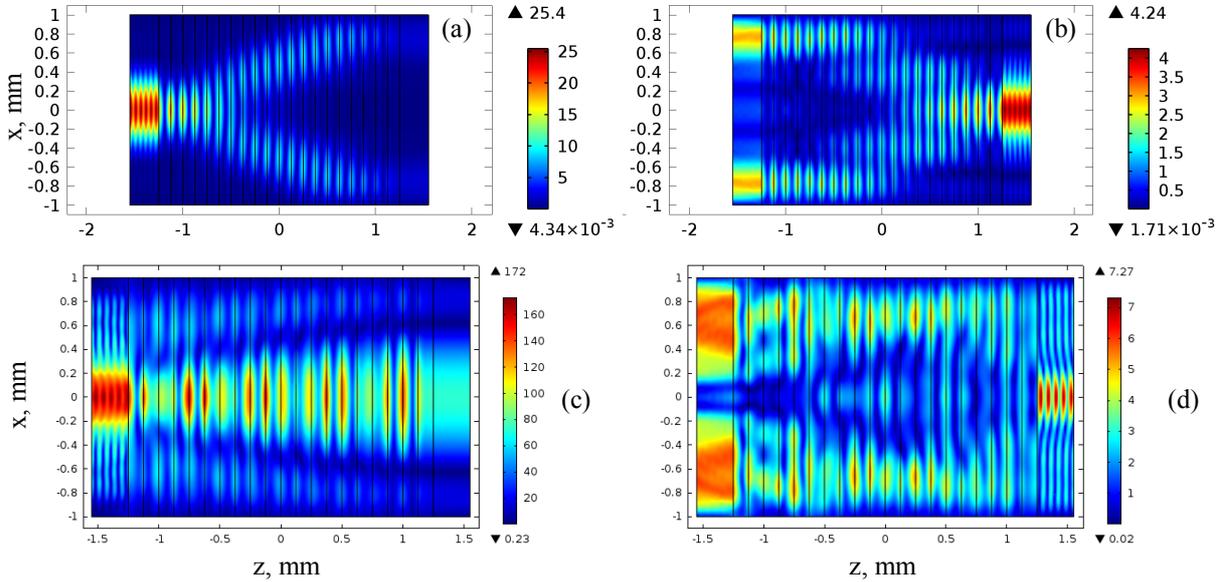

*Figure 4. Distribution of the magnetic-field magnitude $|H_y|$ (the units of measurement are [A/m]) for Gaussian beam normal incidence from left (a,c) and right (b,d) at $\varepsilon' = 0.1$, $\varepsilon'' = 0.1$, $d=125\ \mu m$, $N=10$; (a,b)- $\omega = 2.833\times10^{13}\ s^{-1}$; (c,d)- $\omega = 2.417\times10^{13}\ s^{-1}$.*

The effect of the beam incidence angle on the field distribution is displayed in Fig. 5 at $\theta_i = 5°$, $N=10$ and $\omega = 2.851\times10^{13}\ s^{-1}$ corresponding to the point of ATR. It can be observed that for slant beam incidence the symmetry of field distribution will be broken.

## 5. CONCLUSIONS

The scattering features of *PT*-symmetric periodic stack of binary dielectric layers characterised by balances loss and gain have been explored. The resonant scattering properties and non-reciprocal behaviour of plane wave and Gaussian beam are illustrated by the simulation results. The beam splitting in *PT*-symmetric structures is observed.

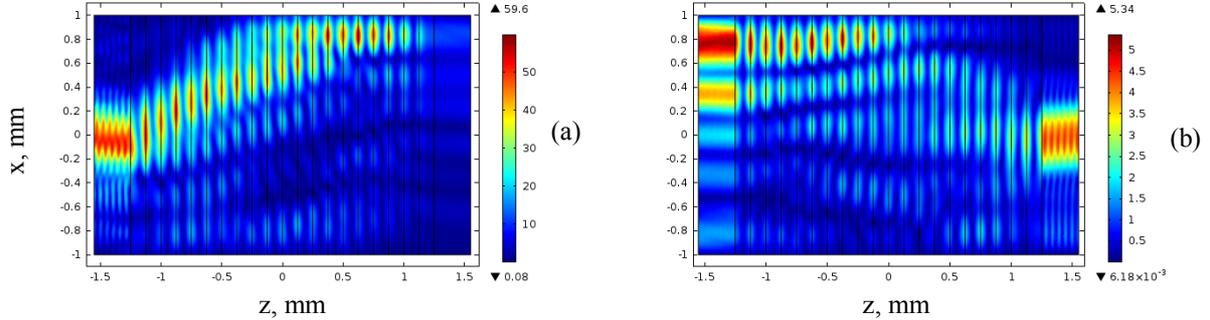

*Figure 5. Distribution of the magnetic-field magnitude $|H_y|$ (the units of measurement are [A/m]) for Gaussian beam incidence from left (a) and right (b) at $\theta_i = 5°$, $\varepsilon' = 0.1$, $\varepsilon'' = 0.1$, $d=125$ μm, $N=10$, $\omega = 2.851 \times 10^{13}$ s$^{-1}$.*

## ACKNOWLEDGEMENTS

The research work was partially supported by the European Union Seventh Framework Program (FP7-REGPOT-2012-2013-1) under grant agreement No. 316165.